\documentclass[aps,prb,showpacs,twocolumn]{revtex4}
\usepackage[utf8x]{inputenc}
\usepackage{amssymb}
\usepackage{graphicx}
\newcommand{\ud}{\mathrm{d}}

\begin{document}
\title{Paramagnetic Effects on the Vortex Lattice of the Unconventional Superconductor CeCoIn$_5$}

\author{V. P. Michal and V. P. Mineev}
\affiliation{Commissariat \`a l'Energie Atomique,
INAC/SPSMS, 38054 Grenoble, France}
\date{\today} 

\begin{abstract}
We expose an analysis of the magnetic field distribution in the Abrikosov lattice of high-$\kappa$ superconductors with d-wave pairing in the case where the critical field is mainly determined by the Pauli limit and the superfluid currents mainly come from the paramagnetic interaction of electron spins with the local magnetic field. The result found in  frame of the generalized Clem variational approach is compatible with the recent observation that the form factor in CeCoIn$_5$ increases with increasing field and then decreases at the approach of $H_{c2}$. 
(A. D. Bianchi et al., Science \textbf{319}, 177 (2008)).\\
\end{abstract}
\maketitle
Recent neutron scattering experiments performed on the heavy-fermion superconductor CeCoIn$_5$
have revealed an unexpected behaviour of the vortex lattice (VL) form factor\cite{Bianchi, White,Esk} defined as the Fourier transform of the local magnetic field in the vortex lattice. The VL form factor of the type II superconductors is usually a decreasing function of the magnetic field.\cite{Clem}
On the contrary, the VL form factor in CeCoIn$_5$ was found to increase with increasing magnetic field and then to fall down at the approach of $H_{c2}$. \cite{Bianchi, White}
CeCoIn$_5$ is a tetragonal, d-wave pairing superconductor with a large GL parameter, and with the highest critical temperature
($T_c=2.3\,K$) among all the heavy fermion compounds.\cite{Petrovic,Izawa,Aoki}
It  has already generated great interest caused by the signs of the existence of the FFLO phase for a magnetic field parallel to the ab plane (and possibly to the c-axis),\cite{Bianchi2} and by the presence of an interval where the superconducting/normal phase transition is first order.\cite{Bianchi3,Miclea}
An explanation of the form factor behaviour has been proposed,\cite{Machida} which was based on a numerical processing of the quasi-classical Eilenberger equations in type II superconductors
with strong paramagnetic contribution.

In this Letter we present an analytic derivation of the magnetic field distribution and VL form factor taking into account the paramagnetic effects. Our analysis is based on the electrodynamic theory of the Abrikosov lattice in the superconducting state where the diamagnetic superfluid currents are mainly determined by the Zeeman interaction of the electron spins with the local magnetic field as developed in the paper\cite{Hou2}.

The orbital and the Zeeman respective contribution is quantified by the Maki parameter $\alpha_M=\sqrt{2}H_{c20}/H_p$, where $H_{c20}=\phi_0/2\pi\xi_0^2$ is the orbital critical field while $H_p=\Delta_0/\sqrt{2}\mu$ is the Pauli limiting field, $\phi_0\simeq 2.07\times 10^{-7}\,G\cdot cm^2$ is the flux quantum. Unlike  
the majority of superconductors, the Pauli limiting field in CeCoIn$_5$ is smaller \cite{Bianchi3} than the orbital critical field by a factor of $> 3$. Hence, the Zeeman interaction plays an important role in the mixed state 
field and current distributions.
We demonstrate analytically how the VL form factor, which decreases with increasing magnetic field in the high temperatures region of the phase diagram, at lower temperatures turns to the behaviour increasing with increasing field.

We shall consider a square VL with wave vector $q=2\pi\sqrt{B/\phi_0}$, formed in a tetragonal
type II superconductor under magnetic field directed along the c-axis. The magnetic induction ${\bf B}=\overline{{\bf h}}$ is determined as the spatial average of the local magnetic field
$\mathbf{h}=\nabla\times\mathbf{A}$.  The GL theory for the VL form factor, valid in the limit $\kappa\gg1$  and for the external field not too close to $H_{c2}$, was developed  by J. Clem. \cite{Clem}
Starting from the general form of the order parameter for an isolated vortex
\begin{equation}
 \Delta(\mathbf{r})=\Delta_\infty f(r) e^{-i\varphi}
 \label{OP}
\end{equation}
($\varphi$ is the angle measured from one of the axes in the ab plane), he proposed to model $f(r)$ by the trial function
\begin{equation}
 f(r)=\frac{r}{R},
 \label{f}
\end{equation}
with $R=\sqrt{r^2+\xi_v^2}$. The variational parameter $\xi_v$ was constrained to minimize the vortex total energy and was found to be in the large $\kappa$ limit $\xi_v=\sqrt{2}\,\xi$, where $\xi$ is the coherence length defined below. He has calculated the field distribution due to the orbital current and  obtained the form factor
\begin{equation}
F_{orb}=B\,\frac{K_1(Q \xi_v)}{Q\lambda K_1(\xi_v/\lambda)},
\label{CFF}
\end{equation}
where  $Q=\sqrt{q^2+\lambda^{-2}}$, $K_1(x)$ is the modified Bessel function of first order,\cite{Abr} and $\lambda$ is the London penetration depth.
One can write an approximative form of it in the conditions $\kappa\gg1$, $q\gg\lambda^{-1}$, and $q\xi_v\ll1$,
\begin{equation}
F_{orb}\simeq \frac{\phi_0}{(2\pi\lambda)^2}-\frac{\xi^2B}{2\lambda^2}\ln\left(\frac{2e\phi_0}{\gamma^2(2\pi\xi)^2B}\right),
\label{Fgamma}
\end{equation}
 $\ln\gamma=C\simeq0.577$ is the Euler constant. 
 
The found form factor, that quite slowly decreases with magnetic field, is reliable if one neglects the paramagnetic interaction of the electron spins with the magnetic field.
The latter leads to two extra features that are important in the case of a large enough Maki parameter.
First, the two characteristic lengths $\xi$ and $\lambda$ in the above expression proves to be magnetic field dependent. Second, a new mechanism originating from the Zeeman interaction gives rise to the main contribution to the diamagnetic screening \cite{Hou2}
in the high magnetic field region of the phase diagram. To find it we start with the Ginzburg-Landau formulation including the paramagnetic effects.

The superconductor CeCoIn$_5$ has pairing symmetry $d_{x^2-y^2}$,\cite{Vorontsov} with order parameter
\begin{equation}
 \Delta_{\mathbf{k}}(\mathbf{r})=\psi(\mathbf{\hat{k}})\Delta(\mathbf{r}),~~~~\psi(\mathbf{\hat{k}})=\sqrt{2}\cos(2\varphi).
\end{equation}
The free energy of the system is given by the Ginzburg-Landau (GL) functional
\begin{eqnarray}
\mathcal{F}=\int\ud^2\mathbf{r}\,\Big(\,\frac{\mathbf{h}^2}{8\pi}+\big(\alpha+\varepsilon(h_z-B)\big)|\Delta|^2\nonumber\\+\beta|\Delta|^4+\gamma|\mathbf{D}\Delta|^2\Big),
\label{GL}
\end{eqnarray}
where $\mathbf{D}=-i\nabla+2e\mathbf{A}$ is the gauge-invariant derivative (from here we put $\hbar=c=1$), and 
the coefficients are functions of temperature and induction. In the clean limit they are\cite{Hou2,Hou1}
\begin{eqnarray}
 \alpha&=&N_0\big(\ln(T/T_c)+\Re\mathfrak{e}\Psi(w)- \Psi (1/2)\big),\nonumber\\
 \varepsilon&=&\frac{N_0\mu}{2\pi T}\Im\mathfrak{m}\Psi'(w),\nonumber\\
 \beta&=&-\frac{N_0}{8(2\pi T)^2}\langle|\psi(\mathbf{\hat{k}})|^4\rangle\Re\mathfrak{e}\Psi^{(2)}(w),\nonumber\\
 \gamma&=&-\frac{N_0 v_F^2}{16(2\pi  T)^2}\Re\mathfrak{e}\Psi^{(2)}(w),\nonumber
\end{eqnarray}
where $\Psi(w)$
is the digamma function, $\Psi^{(m)}(w)$ are its derivatives called by the polygamma functions, \cite{Abr} and 
$$
 w=\frac{1}{2}-\frac{i\mu B}{2\pi  T}\cdot
$$
The coefficient $\varepsilon$ is proportional to $B$. Hence, the corresponding term in the functional is negligibly small in the ordinary GL region near $T_c$ ($B\to 0$).
In a quasi-two-dimensional case we deal in the first approximation with a cylindrical Fermi surface.
For the order parameter given above the average over the Fermi surface is $\langle|\psi(\mathbf{\hat{k}})|^4\rangle=3/2$. 

The stationary condition for the functional in respect of the order parameter gives the non-linear GL equation
\begin{equation}
\alpha_1 \Delta+ 2\beta|\Delta|^2\Delta+\gamma\mathbf{D}^2\Delta=0.
 \label{NonLinear}
\end{equation}
The coherence length is defined as 
$
 \xi=\sqrt{\gamma/|\alpha_1|},
$
with $\alpha_1=\alpha+\varepsilon(h_z-B)$.  
This expression is meaningful away from the region where $\gamma=0$ (close to which the FFLO phase may exist).

The variational solution of the GL equation for an isolated vortex  \cite{Clem} is given by eqns (\ref{OP}), (\ref{f})  with 
the order parameter amplitude
$
 \Delta_\infty=\sqrt{|\alpha_1|/2\beta}.
$
In the high $\kappa$ limit, we have $h\simeq B$ almost everywhere. Therefore as far as the configuration of the order parameter is concerned, we will take $\alpha_1\simeq\alpha$.

The stationary condition with respect to the vector potential  gives the Maxwell  equation 
\begin{equation}
\frac{1}{4\pi}\nabla\times\mathbf{h}={\bf j}_{orb}+{\bf j}_Z
\label{Maxwell}.
\end{equation}
For an isolated vortex with  order parameter given by eqns. (\ref{OP}), (\ref{f}) the vector potential is of the form
$\mathbf{A}_v(\mathbf{r})=A_v(r)\hat{\varphi}$,
the \emph{orbital} density of current is
\begin{equation}
{\bf j}_{orb}=-8e^2\gamma\big(A_v(r)-\frac{\phi_0}{2\pi r}\big)|\Delta|^2\hat{\varphi}
 \label{jgamma},
\end{equation}
while the \emph{Zeeman} current \cite{Hou2} found from (\ref{GL}) is 
\begin{equation}
{\bf j}_{Z}=\varepsilon\frac{d}{dr}|\Delta|^2\hat{\varphi}.
\end{equation}
Hence, we come to the equation that determines the vector potential $A_v(r)$
\begin{equation}
 \frac{d}{dr}\big(\frac{1}{r}\frac{d}{dr}(rA_v)\big)-\frac{f^2}{\lambda^2}A_v=-\frac{\phi_0f^2}{2\pi\lambda^2r}-4\pi\varepsilon\Delta_\infty^2\frac{df^2}{dr},
 \label{DE}
\end{equation}
where $\lambda=\sqrt{\beta/16\pi e^2\gamma|\alpha|}$ is the penetration depth. Writing (\ref{DE}) for the shifted potential $a(r)=A_v(r)-\phi_0/2\pi r$, we obtain the differential equation with an inhomogeneous term of Zeeman origin
\begin{equation}
  \frac{d}{dr}\big(\frac{1}{r}\frac{d}{dr}(ra)\big)-\frac{f^2}{\lambda^2}a=-4\pi\varepsilon\Delta_\infty^2\frac{df^2}{dr}.
\end{equation}
The general solution of this equation $a(r)=a_h(r)+a_i(r)$ consists of the sum of the solutions of the homogeneous and the inhomogeneous equations.
The former determines the orbital part $A_{orb}=\phi_0/2\pi r+a_h$ of the vector potential $A_v=A_{orb}+A_Z$,
\begin{equation}
 A_{orb}(r)=\frac{\phi_0}{2\pi r}\left (1-\frac{R}{\xi_v}\frac{K_1(R/\lambda}{K_1(\xi_v/\lambda}\right ),
 \end{equation}
 such that the corresponding magnetic field ${\bf h}_{orb}=h_{orb}\hat z$ is \cite{Clem}
\begin{equation}
 h_{orb}=\frac{\phi_0}{2\pi\lambda\xi_v}\frac{K_0(R/\lambda)}{K_1(\xi_v/\lambda)},
\end{equation}
and the form factor is determined by eqn.(\ref{CFF}).
 Taking into account the expressions for $\alpha, \beta$ and $\gamma$ we obtain field dependence
for the orbital part of the form factor determined at $\kappa\gg1$ by the  first term in (\ref{Fgamma})
\begin{equation}
 F_{orb}\simeq\frac{4}{3\pi}\phi_0e^2v_F^2|\alpha|.
 \label{Fgammaeq}
\end{equation}
 The Zeeman part of the vector potential is given by
 \begin{equation}
 A_Z(r)=a_i(r)=\frac{R}{r}K_1(R/\lambda) C(R/\lambda)
 \end{equation}
 with
\begin{eqnarray}
 C(z)&=&\int_{\xi_v/\lambda}^z \ud x\,\phi(x), ~~~~~\phi(z)=t(z)e^{-s(z)},\nonumber\\
 t(z)&=&-\frac{8\pi\varepsilon\Delta_\infty^2\xi_v^2}{\lambda}\int \ud z\,\frac{1}{z^3 K_1(z)}e^{s(z)},\nonumber\\
 s(z)&=&-\int \ud z\,\left(\frac{1}{z}+\frac{2K_0(z)}{K_1(z)}\right).\nonumber
\end{eqnarray}
The corresponding magnetic field ${\bf h}_Z=h_Z\hat z$ takes the form
\begin{equation}
 h_Z=\frac{1}{\lambda}\left[K_0(R/\lambda)C(R/\lambda)+K_1(R/\lambda)C^\prime(R/\lambda)\right].
\label{heps}
\end{equation}
In order to find a simpler expression for this new term, we can look at its behaviour in the region not too far from the vortex core i.e. for $R\ll\lambda$, corresponding to the region between the vortices of the lattice. By using the Bessel functions asymptotic expressions in that limit,\cite{Abr} we find the dominating term in the eqn. (\ref{heps})
\begin{equation}
 h_{Z}\simeq4\pi\varepsilon\Delta_\infty^2\frac{\xi_v^2}{R^2}.
\label{hepseq}
\end{equation}
We see that $h_Z(r)$ is concentrated near the core of a vortex, the characteristic length associated to it being of the order of $\xi_v$. 
As a remark, this term may also be derived by considering the equation
$
 \nabla\times\mathbf{h}_Z=4\pi{\bf j}_Z,
$
that is valid in the absence of the orbital current.  From (\ref{heps}) one can find the correction to (\ref{hepseq}),
$
 \delta h_Z=4\pi \varepsilon\Delta_\infty^2K_0(R/\lambda)\ln(R/\xi_v)/\kappa^2$,
which is small in the high $\kappa$ limit. Let us therefore consider expression (\ref{hepseq}) for deriving the new form factor.

The Fourier transform of the magnetic field around a single vortex is
\begin{equation}
{h}_Z(q)= \int \ud^2\mathbf{r}\,h_Z(r)e^{-i{\bf q}\cdot{\bf r}}=8\pi^2\varepsilon \Delta_\infty^2
\xi_v^2K_0(q\xi_v).
\label{Fourier}
\end{equation}
Hence, the contribution to the form factor that originates from the interaction of the electron spins with the local magnetic field for an array of $B/\phi_0$ vortices per $cm^2$ is 
\begin{equation}
 F_{Z}=4\pi\frac{N_0{\mu B}}{\phi_0{T}}\Delta_\infty^2 \Im\mathfrak{m}\Psi'(w)\xi_v^2K_0(q\xi_v)
\end{equation}
One can evaluate the total energy of the magnetic field (quantities integrated over the whole 2-D plane) and deduce that the energy of a single vortex is (in the high $\kappa$ limit) dominated by the term $h_{orb}$.
Therefore the minimization of the energy of a single vortex gives the same variational parameter $\xi_v=\sqrt{2}\,\xi$ as exposed in the Clem paper.\cite{Clem}  As a result we obtain
\begin{equation}
 F_{Z}=\frac{4\pi}{3}\frac{N_0{\mu B}v_F^2}{\phi_0{T}} \Im\mathfrak{m}\Psi'(w)K_0(\sqrt{2}q\xi)
 \label{ZFF}
\end{equation}
\begin{figure}[b]
\centering
\includegraphics[width=9cm]{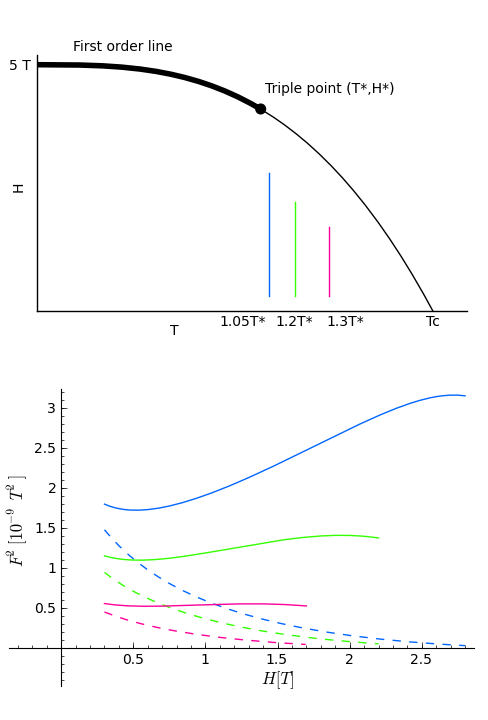}
\caption{(Above) CeCoIn$_5$ phase diagram for $H ||$c-axis. The color lines represent the temperatures where we applied the model. (Below) Variations of the squared form factor $F^2$ at different temperatures including both the orbital and Zeeman contributions. The dashed lines represent the variations of the orbital part alone.}
\label{FormFactor}
\end{figure}
For $q\xi\ll 1$ and $\mu B<2\pi T$
the ratio of two form factors (\ref{ZFF}) and (\ref{Fgammaeq}) is
 \begin{equation}
\frac{F_Z}{F_{orb}}\simeq\frac{7\zeta(3)\mu^2 B^2}{\pi T^2}\frac{\ln(\sqrt{2}/q\xi)}
{\ln(T_c/T)-7\zeta(3)(\mu B/2\pi T)^2},
\end{equation}
here $\zeta(x)$ is the Riemann zeta function. 
We observe that the Zeeman part of the form factor prevails over its orbital part  in the phase diagram region where $\mu B \simeq T$.

We are now able to analyze the behaviour of the total form factor 
\begin{equation}
F=F_{orb}+F_Z,
\end{equation}
where  $F_{orb}$ and $F_Z$ are given by equations (\ref{CFF}) and (\ref{ZFF}) correspondingly.
It is drawn  in Fig. \ref{FormFactor}. In numerical calculations we assumed the values $\mu=g\mu_B/2=\mu_B$ for the electron magnetic moment in the material, and $v_F=5\times 10^5\,cm/s$ for the Fermi velocity inside the superconducting phase. A value for $v_F$ slightly bigger was given in\cite{Miclea} as a result of measurements of the upper critical field $H_{c2}$ near $T_c$. The 2-D density of states on the Fermi surface is independent of $v_F$ and is given by $N_0=m^*/2\pi\ell_c$, where we considered $m^*=100\,m_e$ for the electron effective mass, and $\ell_c=7.6\times10^{-8}\,cm$ is the lattice c-axis spacing.

In the form factor variations, there is first a domination of the orbital part in the low magnetic field region ($F_Z$ vanishes at $B=0$). We observe next a crossover to a region where the paramagnetic term is dominant. The regime where (\ref{CFF}) goes exactly against (\ref{ZFF}) is likely to explain the observed constant logarithm of the squared form factor\cite{Esk} in the interval $B=0.5-2T$.  In addition, these features were observed in an experiment realised on the s-wave superconductor TmNi$_2$B$_2$C.\cite{Esk2} At larger fields and smaller temperatures region, the form factor increases with field. 

The observed form factor\cite{Bianchi,White} falling towards zero near the phase transition  line,
where the field variation amplitude decreases with the order parameter,
 is out of the region of applicability of the developed theory where isolated vortices are supposed. To describe the form factor behaviour reaching the maximum and decreasing at the approach of the upper critical field, one must match the present theory to the Ginzburg-Landau description valid in the vicinity of $H_{c2}$
at temperatures above and also slightly below the tricritical point where the order parameter $\Delta_\infty^2$ takes a finite value at the critical field.
This will be published elsewhere.

In conclusion, making use of the generalized Clem approach 
 we have calculated the magnetic field dependence of the vortex lattice form factor.
Some results found previously in the framework of the Ginzburg-Landau-Abrikosov theory\cite{Hou2} lead us to find a new term originating from the interaction of the electron spins with the magnetic field existing inside the sample. The latter has a dominant contribution in the expression of the vortex lattice form factor at high magnetic field in the superconductors with a small enough Fermi velocity.\cite{footnote}
The magnetic field caused by the diamagnetic currents originating from the Zeeman interaction was found to be concentrated in the core of vortices and gives rise to new features of the form factor that accounts for the measurements that are currently being made on CeCoIn$_5$.\cite{Bianchi,White,Esk}

\end{document}